\documentclass{aa}

\usepackage{graphicx}
\usepackage{amssymb}
\usepackage{natbib}
\usepackage{array}
\usepackage{amsmath}
\usepackage{subfigure}
\usepackage{color}
\usepackage{longtable}
\bibpunct{(}{)}{;}{a}{}{,}

\begin{document}

\title{Factors Affecting the Radii of Close-in Transiting Exoplanets}
\subtitle{}

\author{B.Enoch\inst{1}
\and A.Collier Cameron\inst{1}
\and K.Horne\inst{1}
}

\institute{School of Physics and Astronomy, University of St. Andrews, North Haugh, St Andrews, KY16 9SS}

\authorrunning{}

\date{Received / Accepted}

\abstract
{The radius of an exoplanet may be affected by various factors, including irradiation received from the host star, the mass of the planet and its heavy element content. A significant number of transiting exoplanets have now been discovered for which the mass, radius, semi-major axis, host star metallicity and stellar effective temperature are known.}
{We use multivariate regression models to determine the power-law dependence of planetary radius on planetary equilibrium temperature $T_{eq}$, planetary mass $M_p$, stellar metallicity [Fe/H], orbital semi-major axis $a$, and tidal heating rate $H_{tidal}$, for 119 transiting planets in three distinct mass regimes.}
{We fit models initially to all 119 planets, resulting in fairly high scatter between fitted and observed radii, and subsequently to three subsets of these planets: Saturn-mass planets, Jupiter-mass planets, and high-mass planets.}
{We find models for each subset that fit the observed planetary radii well and show the importance of the various environmental parameters on each subset.}
{We determine that heating leads to larger planet radii, as expected, increasing mass leads to increased or decreased radii of low-mass ($<0.5 R_J$) and high-mass ($>2.0 R_J$) planets, respectively (with no mass effect on Jupiter-mass planets), and increased host-star metallicity leads to smaller planetary radii, indicating a relationship between host-star metallicity and planet heavy element content. For Saturn-mass planets, a good fit to the radii may be obtained from log($R_p/R_J$) = -0.077 + 0.450 log($M_p/M_J$) - 0.314 $[$Fe/H$]$ + 0.671 log($a$/AU) + 0.398 log($T_{eq}$/K). The radii of Jupiter-mass planets may be fit by log($R_p/R_J$) = -2.217 + 0.856 log($T_{eq}/K$) + 0.291 log($a/AU$). High-mass planets' radii are best fit by log($R_p/R_J$) = -1.067 + 0.380 log($T_{eq}/K$) - 0.093 log($M_p/M_J$) - 0.057 $[$Fe/H$]$ + 0.019 log($H_{tidal}/1\times10^{20}$). These equations produce a very good fit to the observed radii, with a mean absolute difference between fitted and observed radius of 0.11~$R_J$, compared to the mean reported uncertainty in observed radius of 0.07~$R_J$. A clear distinction is seen between the core-dominated Saturn-mass (0.1-0.5~$M_J$) planets, whose radii are determined almost exclusively by their mass and heavy element content, and the gaseous envelope-dominated Jupiter-mass (0.5-2.0~$M_J$) planets, whose radii increase strongly with irradiating flux, partially offset by a power-law dependence on orbital separation.}


\maketitle

\section{Introduction}
\label{s:intro}

The mass-radius relationship of gravitationally bound objects may be approximated by a polytropic relationship given by $R \propto M^{(1-n)/(3-n)}$  \citep{burrows93}. The polytropic index $n$ ranges from 0 for low mass planets made of incompressible matter, i.e. $R \propto M^{1/3}$, to 3/2 for low-mass stars with electron degeneracy, i.e. $R \propto M^{-1/3}$ \citep{chabrier09}. Between these two regimes, at approximately Jupiter-mass objects \citep{zapolsky69}, $n \sim 1$, i.e. $R \propto M^0$, showing that the radii of Jupiter-mass planets do not depend on their masses. The variation in radii of most known exoplanets must therefore arise from other influencing factors. For example, \citet{guillot05} derive a mass-radius relationship for highly irradiated planets, incorporating heating effects. 

Various potential effects on planetary radii have been discussed in the literature. The equilibrium temperature of the planet, related to the irradiating flux received from the host star, is likely to affect the radius \citep{guillot96,guillot02} in that planets that are strongly heated by their host stars may have inflated radii. 

Additional heating effects have also been proposed, including Ohmic heating \citep{batygin10,batygin11,laughlin11}, a mechanism of planetary heating from the coupling of the planetary magnetic field and atmospheric flows; kinetic heating \citep{guillot02} where a small amount of flux from the host star is transformed into kinetic energy and then thermal energy in the atmosphere of the planet; and tidal heating \citep{bodenheimer01,bodenheimer03,jackson08} due to circularisation of close-in exoplanet orbits.

The metallicity of the system is also thought to affect planetary radii. Increasing the metallicity could lead to an increased heavy element abundance in the atmosphere, producing higher atmospheric opacity which could again retard the loss of heat from a planet and thus slow its contractions \citep{burrows07}. However, an enriched atmosphere would also have a higher mean molecular weight, offsetting partially or totally any reduction in contraction \citep{burrows07, guillot08}. Alternatively, the higher metallicity may result in increased heavy elements in the core, leading to a decreased planetary radius relative to a planet of the same mass but lower metallicity \citep{guillot06}. \citet{fortney09} set this out in their `basic questions' about planetary structure and composition, including whether heavy elements in a planet mix into the H-He envelope, or are found in the core, and whether a planet's heavy element mass fraction depends on that of its parent star.

Recently, \citet{laughlin11} found a correlation between the radius anomalies from comparing observed radii with model radii based on calculations of coreless giant planets tabulated for various mass and insolation values \citep{bodenheimer03}. They found an additional radius dependence on planetary temperature of $R \propto T^{1.4}$, and suggested Ohmic heating may be largely responsible. They also noted that even including this additional dependence, there remained significant scatter in the observed radii and suggested evidence of a signficant correlation with host star metallicities, and that tidal heating may also contribute.

The system age may also have an effect since the radius of an isolated planet should decrease with time due to Kelvin-Helmholtz contraction. 

Additionally, \citet{valencia10} and \citet{jackson10} recently investigated CoRoT-7b, concluding that it is likely to have initially had a much larger mass than currently, possibly including a substantial gaseous component which would have given it a larger radius than the 0.15~$R_J$ observed today. Close-in planets are susceptible to evaporation due to intense radiation from the star. A large loss of mass (and radius) can occur due to enhanced atmospheric escape for exoplanets that are very close to their stars. Molecules are able to escape even if below the normal escape velocity, needing only to reach the Roche Lobe. The atmosphere of H-He planets is only loosely bound to the planetary core, so evaporation can be quicker than the contraction of a planet, which may result in loss of the whole envelope for a planet of up to $1 M_J$ initially \citep{valencia10}. A planet which lost a significant fraction of mass may become stable when the EUV flux from the young host star drops \citep{valencia10}. Models by \citet{hubbard07} show significant initial mass loss in the first 0.1~Gyr of a system. \citet{baraffe05} show that recent Neptunian objects could have originally been more massive irradiated gas giants.

The reduction in required escape energy may be given by \citep{jackson10,erkaev07} 
\begin{equation}
K_{tide} = 1 - \frac{3}{2\xi} + \frac{1}{2 \xi^3}
\label{ktide}
\end{equation}
\noindent where
\begin{equation}
\xi = \frac{(M_p / 3 M_{\ast})^{1/3} a}{R_p}
\end{equation}
\noindent The mass loss rate due to stellar irradiation is then given by 
\begin{equation}
\frac{dM_p}{dt} = - \frac{\pi R_p^3 \epsilon F_{xuv}}{GM_pK_{tide}}
\end{equation}
\noindent where $F_{xuv}$ is the extreme UV flux from the star and $\epsilon$ is the fraction of incoming energy that is removed by escaping gas. Thus planets of smaller semi-major axis may be trimmed down by atmospheric blow-off. Atmospheric escape of HD~209458, with $K_{tide} = 0.65$, has been observed in the Ly~$\alpha$ line \citep{vidal03}. More recently, \citet{fossati10} report finding enhanced transit depths when observing the very close-in exoplanet WASP-12b with the HST/COS instrument using UV transmission spectroscopy. All the planets considered here have $K_{tide}$ values below 1, while some have very low values and thus significant enhancement of mass loss, for example WASP-19b and WASP-12b with $K_{tide}$ values of 0.29 and 0.27 respectively.

We previously investigated the effect of metallicity and equilibrium temperature on the radii of planets in a mass range of $0.1-0.6 M_J$, finding a strong correlation of stellar radius with irradiating flux, and a weaker anticorrelation with host-star metallicity \citep{enoch10}. In this paper, we seek to illuminate the main trends seen in the radii of the majority of known transiting exoplanets with a wide range of masses and radii using several parameters. Section \ref{s:analysis} sets out the method of analysis and gives the resulting calibrations for the planetary radii, while Section \ref{s:disc} provides a discussion of the results. 

\section{Analysis}
\label{s:analysis}

There are currently (November 2011) almost 200 known transiting exoplanets\footnotemark \footnotetext[1]{www.exoplanet.eu}, for which values have been published for both mass and radius. To perform our analysis we selected all known transiting planets in the mass range 0.1 - 12.0~$M_J$ with orbital periods below 10 days for which values have been published for host star metallicity and effective temperature. This resulted in a sample of 119 planets, given in Tables \ref{tab:satplanets} to \ref{tab:highplanets}. 

\begin{table*}[h!]
\begin{center}
\setlength{\extrarowheight}{3pt}
\caption{Saturn-mass transiting planets used in the analysis of radii.}
\label{tab:satplanets}
\begin{tabular}{lrrrrrrrr}
\hline
ID & $M_p$ ($M_J$) & $R_p$ ($R_J$) & $a$ (AU) & $e$ & $T_{eff}$ (K) & $[$Fe/H$]$ & $R_{\ast}$ ($R_{\odot}$) & Age (Gy) \\
\hline
HAT-18b & $0.20\pm0.01$ & $1.00\pm0.05$ & $0.0559\pm0.0007$ & 0 & $4803\pm80$ & $0.10\pm0.08$ & $0.75\pm0.04$ & 12.4 \\
HAT-12b & $0.21\pm0.01$ & $0.96^{+0.03}_{-0.02}$ & $0.0384\pm0.0003$ & 0 & $4650\pm60$ & $-0.29\pm0.05$ & $0.70^{+0.02}_{-0.01}$ & 2.0 \\
CoRoT-8b & $0.22\pm0.03$ & $0.57\pm0.02$ & $0.063\pm0.001$ & 0 & $5080\pm80$ & $0.3\pm0.1$ & $0.77\pm0.02$ & 2.0 \\
WASP-29b & $0.25\pm0.02$ & $0.74\pm0.06$ & $0.0456\pm0.0006$ & 0 & $4800\pm150$ & $0.11\pm0.14$ & $0.85\pm0.05$ & 5.0 \\
WASP-39b & $0.28\pm0.03$ &  $1.27\pm0.04$ & $0.0486\pm0.0005$ & 0 & $5400\pm150$ & $-0.12\pm0.1$ & $0.90\pm0.02$ & 9.0 \\
HAT-19b & $0.29\pm0.02$ & $1.13\pm0.07$ & $0.0466\pm0.0008$ & 0 & $4990\pm130$ & $0.23\pm0.08$ & $0.82\pm0.05$ & 8.8 \\
WASP-21b & $0.30\pm0.01$ & $1.07\pm0.05$ & $0.0520\pm0.0004$ & 0 & $5800\pm100$ & $-0.4\pm0.1$ & $1.06\pm0.04$ & 5.0 \\
HD149026b & $0.37\pm0.01$ & $0.81\pm0.03$ & $0.0431^{+0.0007}_{-0.0006}$ & 0 & $6147\pm50$ & $0.36\pm0.05$ & $1.54^{+0.05}_{-0.04}$ & 0.2 \\
Kepler-7b & $0.43\pm0.04$ & $1.48\pm0.05$ & $0.0622^{+0.0011}_{-0.0008}$ & 0 & $5933\pm44$ & $0.11\pm0.03$ & $1.84\pm0.07$ & 1.0 \\
Kepler-12b & $0.43\pm0.04$ & $1.70\pm0.03$ & $0.0556\pm0.0007$ & 0 & $5947\pm100$ & $0.07\pm0.04$ & $1.48\pm0.03$ & 4.0 \\
WASP-11b & $0.46\pm0.03$ & $1.05^{+0.05}_{-0.03}$ & $0.043\pm0.002$ & 0 & $4980\pm60$ & $0.13\pm0.08$ & $0.81\pm0.03$ & 4.1 \\
WASP-13b & $0.46^{+0.06}_{-0.05}$ & $1.21\pm0.14$ & $0.0527^{+0.0017}_{-0.0019}$ & 0 & $5826\pm100$ & $0.0\pm0.2$ & $1.34\pm0.13$ & 4.9 \\
CoRoT-5b & $0.47^{+0.07}_{-0.02}$ & $1.39^{+0.04}_{-0.05}$ & $0.0495\pm0.0003$ & 0 & $6100\pm65$ & $-0.25\pm0.06$ & $1.19\pm0.04$ & 1.4 \\
WASP-31b & $0.48\pm0.03$ & $1.54\pm0.06$ & $0.0466\pm0.0003$ & 0 & $6200\pm100$ & $-0.19\pm0.09$ & $1.24\pm0.04$ & 4.0 \\
WASP-17b & $0.49\pm0.03$ & $1.99\pm0.08$ & $0.0515\pm0.0034$ & 0.03 & $6650\pm80$ & $-0.19\pm0.09$ & $1.57\pm0.06$ & 2.7 \\
WASP-6b & $0.50^{+0.02}_{-0.04}$ & $1.22\pm0.05$ & $0.0421^{+0.0008}_{-0.0013}$ & 0 & $5450\pm100$ & $-0.20\pm0.09$ & $0.87^{+0.03}_{-0.04}$ & 7.0 \\
\hline
\end{tabular}
\end{center}
\end{table*}

\begin{table*}[h!]
\begin{center}
\setlength{\extrarowheight}{3pt}
\caption{Jupiter-mass transiting planets used in the analysis of radii (part 1).} 
\label{tab:jupplanets}
\begin{tabular}{lrrrrrrrr}
\hline
ID & $M_p$ ($M_J$) & $R_p$ ($R_J$) & $a$ (AU) & $e$ & $T_{eff}$ (K) & $[$Fe/H$]$ & $R_{\ast}$ ($R_{\odot}$) & Age (Gy) \\
\hline 
HAT-1b & $0.52\pm0.03$ & $1.23\pm0.06$ & $0.0553\pm0.0014$ & 0 & $6047\pm56$ & ~$0.12\pm0.05$ & $1.12\pm0.05$ & 1.0 \\
OGLE-111b & $0.53\pm0.11$ & $1.07\pm0.05$ & $0.047\pm0.001$ & 0 & $5070\pm400$ & ~$0.12\pm0.28$ & $0.83\pm0.03$ & - \\
HAT-17b & $0.53\pm0.05$ & $1.01\pm0.05$ & $0.0882\pm0.0010$ & 0.35 & $5246\pm100$ & ~$0.0\pm0.1$ & $0.84\pm0.05$ & 7.8 \\
WASP-15b & $0.54\pm0.05$ & $1.43\pm0.08$ & $0.0499\pm0.0018$ & 0 & $6300\pm100$ & $-0.17\pm0.11$ & $1.48\pm0.07$ & 1.3 \\
CoRoT-16b & $0.54\pm0.09$ & $1.17\pm0.15$ & $0.0618\pm0.0015$ & 0.33 & $5650\pm100$ & $0.19\pm0.06$ & $1.19\pm0.14$ & 6.73 \\
WASP-22b & $0.56\pm0.02$ & $1.12\pm0.04$ & $0.0468\pm0.0004$ & 0 & $6000\pm100$ & $-0.05\pm0.10$ & $1.13\pm0.03$ & - \\
XO-2b & $0.57\pm0.06$ & $0.97\pm0.03$ & $0.0369\pm0.002$ & 0 & $5340\pm32$ & ~$0.45\pm0.02$ & $0.96\pm0.02$ & 0.7 \\
HAT-25b & $0.57\pm0.05$ & $1.19\pm0.05$ & $0.047\pm0.001$ & 0 & $5500\pm100$ & ~$0.31\pm0.10$ & $0.96\pm0.05$ & 3.2 \\
WASP-25b & $0.58\pm0.04$ & $1.22^{+0.06}_{-0.05}$ & $0.0473\pm0.0004$ & 0 & $5703\pm100$ & $-0.07\pm0.10$ & $0.92\pm0.04$ & 0.1 \\
WASP-34b & $0.59\pm0.01$ & $1.22^{+0.11}_{-0.08}$ & $0.0524\pm0.0004$ & 0.04 & $5700\pm100$ & $-0.02\pm0.1$ & $0.93\pm0.12$ & 6.7 \\
HAT-3b & $0.60\pm0.03$ & $0.89\pm0.05$ & $0.0389\pm0.0007$ & 0 & $5185\pm46$ & ~$0.27\pm0.04$ & $0.82\pm0.04$ & 0.3 \\
Kepler-8b & $0.60^{+0.13}_{-0.19}$ & $1.42\pm0.06$ & $0.0483^{+0.0006}_{-0.0012}$ & 0 & $6213\pm150$ & $-0.06\pm0.03$ & $1.49\pm0.06$ & 1.5 \\
TrES-1b & $0.61\pm0.06$ & $1.08\pm0.03$ & $0.0393\pm0.0007$ & 0 & $5250\pm200$ & ~$0.0\pm0.2$ & $0.82\pm0.02$ & 1.4 \\
OGLE-10b & $0.63\pm0.14$ & $1.26\pm0.07$ & $0.0416\pm0.0001$ & 0 & $5800\pm100$ & ~$0.0\pm0.2$ & $1.16\pm0.06$ & - \\
HAT-28b & $0.63\pm0.04$ & $1.212^{+0.11}_{-0.08}$  & $0.0434\pm0.0007$ & 0 & $5680\pm90$ & $0.12\pm0.08$ & $1.10^{+0.09}_{-0.07}$ & 6.1 \\
HAT-27/WASP-40b & $0.66\pm0.03$ & $1.06^{+0.05}_{-0.04}$ & $0.0403\pm0.0005$ & 0 & $5300\pm90$ & $0.29\pm0.10$ & $0.90^{+0.05}_{-0.04}$ & 4.4 \\
Kepler-15b & $0.66\pm0.09$ & $0.96\pm0.07$ & $0.0571\pm0.0009$ & 0 & $5595\pm120$ & $0.36\pm0.07$ & $0.99\pm0.07$ & 3.7 \\
Kepler-6b & $0.67\pm0.03$ & $1.32\pm0.03$ & $0.0457^{+0.0006}_{-0.0005}$ & 0 & $5647\pm44$ & ~$0.34\pm0.04$ & $1.39^{+0.02}_{-0.03}$ & 1.0 \\
HAT-4b & $0.68\pm0.04$ & $1.27\pm0.05$ & $0.0446\pm0.0012$ & 0 & $5860\pm80$ & ~$0.24\pm0.08$ & $1.59\pm0.07$ & 0.6 \\
HAT-24b & $0.69\pm0.05$ & $1.24\pm0.05$ & $0.0465\pm0.001$ & 0.07 & $6373\pm100$ & $-0.16\pm0.10$ & $1.32\pm0.05$ & 2.8 \\
HD209458b & $0.69\pm0.02$ & $1.32^{+0.02}_{-0.03}$ & $0.0471\pm0.0005$ & 0 & $6000\pm50$ & ~$0.00\pm0.02$ & $1.15\pm0.06$ & 2.0 \\
HAT-30/WASP-51b & $0.71\pm0.03$ & $1.34\pm0.07$ & $0.0419\pm0.0005$ & 0 & $6250\pm100$ & $-0.08\pm0.08$ & $1.33\pm0.03$ & 1.1 \\
CoRoT-4b & $0.72\pm0.08$ & $1.19^{+0.06}_{-0.05}$ & $0.090\pm0.001$ & 0 & $6190\pm60$ & ~$0.00\pm0.15$ & $1.15^{+0.01}_{-0.03}$ & 0.3 \\
WASP-35b & $0.72\pm0.06$ & $1.32\pm0.03$ & $0.0432\pm0.0003$ & 0 & $6050\pm100$ & $-0.15\pm0.09$ & $1.09\pm0.02$ & 5.0 \\
HAT-33b & $0.76\pm0.12$ & $1.83\pm0.29$ &  $0.0503\pm0.0011$ & 0 & $6401\pm88$ & $0.05\pm0.08$ & $1.78\pm0.28$ & 2.4 \\
HAT-9b & $0.78\pm0.09$ & $1.40\pm0.06$ & $0.053\pm0.002$ & 0 & $6350\pm150$ & ~$0.12\pm0.20$ & $1.32\pm0.07$ & 1.4 \\
HAT-29b & $0.78^{+0.04}_{-0.08}$ & $1.11^{+0.14}_{-0.08}$ & $0.0667\pm0.0008$ & 0 & $6087\pm88$ & $0.21\pm0.08$ & $1.22^{+0.13}_{-0.08}$ & 2.2 \\
TrES-4b & $0.84\pm0.07$ & $1.67\pm0.06$ & $0.0509\pm0.0007$ & 0 & $6100\pm150$ & ~$0.00\pm0.09$ & $1.74\pm0.09$ & 2.0 \\
HAT-13b & $0.85\pm0.04$ & $1.28\pm0.08$ & $0.0426^{+0.0006}_{-0.0012}$ & 0 & $5638\pm90$ & ~$0.43\pm0.08$ & $1.56\pm0.08$ & 0.8 \\
WASP-16b & $0.86\pm0.06$ & $1.01\pm0.07$ & $0.0421^{+0.0010}_{-0.0020}$ & 0 & $5550\pm130$ & ~$0.01\pm0.10$ & $0.95\pm0.05$ & 2.2 \\
WASP-23b & $0.88\pm0.10$ & $0.96\pm0.06$ & $0.0376^{+0.0016}_{-0.0024}$ & 0 & $5150\pm100$ & $-0.05\pm0.13$ & $0.77^{+0.03}_{-0.05}$ & - \\
WASP-1b & $0.89\pm0.20$ & $1.36\pm0.10$ & $0.0382\pm0.0013$ & 0 & $6200\pm200$ & ~$0.26\pm0.03$ & $1.38\pm0.10$ & 1.0 \\
WASP-44b & $0.89\pm0.06$ & $1.14\pm0.11$ & $0.0347\pm0.0004$ & 0 & $5410\pm150$ & $0.06\pm0.10$ & $0.93\pm0.07$ & 0.9 \\
XO-1b & $0.90\pm0.07$ & $1.18\pm0.04$ & $0.0488\pm0.0005$ & 0 & $5750\pm13$ & ~$0.02\pm0.04$ & $0.93\pm0.03$ & 2.0 \\
WASP-28b & $0.91\pm0.06$ & $1.12\pm0.06$ & $0.046\pm0.0005$ & 0 & $6100\pm150$ & $-0.29\pm0.10$ & $1.05\pm0.06$ & 2.0 \\
WASP-2b & $0.91\pm0.09$ & $1.02\pm0.01$ & $0.0307\pm0.0011$ & 0 & $5200\pm200$ & $-0.08\pm0.08$ & $0.83\pm0.08$ & - \\
WASP-41b & $0.92\pm0.07$ & $1.21\pm0.07$ & $0.0400\pm0.0005$ & 0 & $5450\pm150$ & $-0.08\pm0.09$ & $1.01\pm0.26$ & 1.8 \\
CoRoT-12b & $0.92\pm0.07$ & $1.44\pm0.13$ & $0.0402\pm0.0009$ & 0 & $5675\pm80$ & ~$0.16\pm0.10$ & $1.12\pm0.09$ & 6.3 \\
HAT-32b & $0.94\pm0.17$ & $2.04\pm0.10$ & $0.0344^{+0.0004}_{-0.0007}$ & 0 & $6001\pm88$ & $-0.16\pm0.08$ & $1.39\pm0.07$ & 3.8 \\
WASP-7b & $0.96\pm0.13$ & $1.33\pm0.09$ & $0.0617\pm0.0010$ & 0 & $6400\pm100$ & ~$0.0\pm0.1$ & $1.43\pm0.09$ & 2.4 \\
WASP-48b & $0.98\pm0.09$ & $1.67\pm0.08$  & $0.0344\pm0.0004$ & 0 & $5990\pm90$ & $-0.12\pm0.12$ & $1.75\pm0.09$ & 7.9 \\
OGLE-182b & $1.01\pm0.15$ & $1.13^{+0.13}_{-0.08}$ & $0.051\pm0.001$ & 0 & $5924\pm64$ & ~$0.37\pm0.08$ & $1.14^{+0.23}_{-0.06}$ & - \\
WASP-45b & $1.01\pm0.05$ & $1.16^{+0.28}_{-0.14}$ & $0.0405\pm0.0009$ & 0 & $5140\pm200$ & $0.36\pm0.12$ & $0.95^{+0.09}_{-0.07}$ &  1.4 \\
WASP-26b & $1.02\pm0.03$ & $1.32\pm0.08$ & $0.040\pm0.003$ & 0 & $5950\pm100$ & $-0.02\pm0.09$ & $1.34\pm0.06$ & 2.0 \\
CoRoT-1b & $1.03\pm0.12$ & $1.49\pm0.08$ & $0.0254\pm0.0004$ & 0 & $5950\pm150$ & $-0.30\pm0.25$ & $1.11\pm0.05$ & - \\
OGLE-211b & $1.03\pm0.20$ & $1.36^{+0.18}_{-0.09}$ & $0.051\pm0.001$ & 0 & $6325\pm91$ & ~$0.11\pm0.10$ & $1.64^{+0.21}_{-0.07}$ & - \\
WASP-24b & $1.03\pm0.04$ & $1.10^{+0.05}_{-0.06}$ & $0.0359\pm0.0003$ & 0 & $6075\pm100$ & ~$0.07\pm0.10$ & $1.15^{+0.04}_{-0.05}$ & 1.6 \\
\hline
\end{tabular}
\end{center}
\end{table*}

\begin{table*}[h!]
\begin{center}
\setlength{\extrarowheight}{3pt}
\caption{(continued) Jupiter-mass transiting planets used in the analysis of radii.} 
\label{tab:jupplanets2}
\begin{tabular}{lrrrrrrrr}
\hline
ID & $M_p$ ($M_J$) & $R_p$ ($R_J$) & $a$ (AU) & $e$ & $T_{eff}$ (K) & $[$Fe/H$]$ & $R_{\ast}$ ($R_{\odot}$) & Age (Gy) \\
\hline 
HAT-6b & $1.06\pm0.12$ & $1.33\pm0.06$ & $0.0524\pm0.0009$ & 0 & $6570\pm80$ & $-0.13\pm0.08$ & $1.46\pm0.06$ & 0.7 \\
HAT-5b & $1.06\pm0.11$ & $1.26\pm0.05$ & $0.0408\pm0.0008$ & 0 & $5960\pm100$ & ~$0.24\pm0.15$ & $1.17\pm0.05$ & 1.8 \\
XO-5b & $1.08\pm0.04$ & $1.09\pm0.06$ & $0.0487\pm0.0006$ & 0 & $5510\pm44$ & ~$0.18\pm0.03$ & $1.06\pm0.05$ & 0.8 \\
Qatar-1b & $1.09\pm0.08$ & $1.16\pm0.05$ & $0.0234\pm0.0003$ & 0 & $4861\pm125$ & $0.2\pm0.1$ & $0.82\pm0.03$ & 4.0 \\
CoRoT-19b & $1.11\pm0.06$ & $1.45\pm0.05$ & $0.0518\pm0.0008$ & 0 & $6090\pm70$ & $-0.02\pm0.1$ & $1.65\pm0.04$ & 5.0 \\
WASP-4b & $1.12\pm0.09$ & $1.42^{+0.04}_{-0.07}$ & $0.023\pm0.001$ & 0 & $5500\pm150$ & ~$0.0\pm0.2$ & $1.15\pm0.28$ & - \\
HD189733b & $1.13\pm0.03$ & $1.14\pm0.03$ & $0.0310\pm0.0006$ & 0 & $4980\pm200$ & $-0.03\pm0.04$ & $0.79\pm0.05$ & - \\
OGLE-132b & $1.14\pm0.12$ & $1.18\pm0.07$ & $0.0306\pm0.0008$ & 0 & $6210\pm59$ & ~$0.37\pm0.07$ & $1.34\pm0.08$ & - \\
WASP-19b & $1.15\pm0.08$ & $1.31\pm0.06$ & $0.0164^{+0.0005}_{-0.0006}$ & 0 & $5500\pm100$ & ~$0.02\pm0.09$ & $0.93^{+0.05}_{-0.04}$ & 4.5 \\
TrES-2b & $1.20\pm0.05$ & $1.27\pm0.04$ & $0.0356\pm0.0008$ & 0 & $5850\pm50$ & $-0.15\pm0.10$ & $1.00\pm0.04$ & 2.7 \\
OGLE-56b & $1.29\pm0.12$ & $1.30\pm0.05$ & $0.0225\pm0.0004$ & 0 & $6119\pm62$ & ~$0.25\pm0.08$ & $1.32\pm0.06$ & - \\
CoRoT-13b & $1.31\pm0.07$ & $1.25\pm0.08$ & $0.051\pm0.003$ & 0 & $5945\pm90$ & ~$0.01\pm0.07$ & $1.01\pm0.03$ & 1.6 \\
OGLE-113b & $1.32\pm0.19$ & $1.09\pm0.03$ & $0.0229\pm0.0002$ & 0 & $4804\pm106$ & ~$0.14\pm0.02$ & $0.77\pm0.02$ & - \\
WASP-12b & $1.41\pm0.10$ & $1.79\pm0.09$ & $0.0229\pm0.0008$ & 0 & $6300\pm150$ & ~$0.3\pm0.1$ & $1.57\pm0.07$ & 0.8 \\
WASP-50b & $1.47\pm0.09$ & $1.15\pm0.05$ & $0.0295\pm0.0009$ & 0 & $5400\pm100$ & $-0.12\pm0.08$ & $0.84\pm0.03$ & 7.0 \\
HAT-8b & $1.52^{+0.18}_{-0.16}$ & $1.50^{+0.08}_{-0.06}$ & $0.0487\pm0.0026$ & 0 & $6200\pm80$ & ~$0.01\pm0.08$ & $1.58^{+0.08}_{-0.06}$ & 1.0 \\
WASP-5b & $1.64\pm0.08$ & $1.17\pm0.06$ & $0.0273\pm0.0006$ & 0 & $5880\pm150$ & ~$0.0\pm0.2$ & $1.08\pm0.04$ & 1.4 \\
XO-4b & $1.72\pm0.20$ & $1.34\pm0.05$ & $0.0555\pm0.0011$ & 0 & $5700\pm70$ & $-0.04\pm0.03$ & $1.55\pm0.05$ & 0.6 \\
WASP-3b & $1.76^{+0.06}_{-0.14}$ & $1.31^{+0.05}_{-0.12}$ & $0.0317^{+0.0006}_{-0.001}$ & 0 & $6400\pm100$ & ~$0.0\pm0.2$ & $1.31^{+0.06}_{-0.12}$ & 1.4 \\
TrES-5b & $1.78\pm0.06$ & $1.21\pm0.02$ & $0.0245\pm0.0001$ & 0 & $5171\pm36$ & $0.20\pm0.08$ & $0.87\pm0.01$ & 7.38 \\
WASP-43b & $1.78\pm0.10$ & $0.93^{+0.07}_{-0.09}$ & $0.0142\pm0.0004$ & 0 & $4400\pm200$ & $-0.05\pm0.17$ & $0.60\pm0.04$ & 0.4 \\
HAT-7b & $1.80\pm0.06$ & $1.42^{+0.14}_{-0.10}$ & $0.0379\pm0.0004$ & 0 & $6350\pm80$ & ~$0.26\pm0.08$ & $1.84^{+0.23}_{-0.11}$ & 1.0 \\
WASP-37b & $1.80\pm0.17$ & $1.16^{+0.07}_{-0.06}$ & $0.0446\pm0.0019$ & 0 & $5800\pm150$ & $-0.40\pm0.12$ & $1.00\pm0.05$ & 11 \\
TrES-3b & $1.92\pm0.23$ & $1.30\pm0.08$ & $0.0226\pm0.0013$ & 0 & $5720\pm150$ & $-0.19\pm0.08$ & $0.81^{+0.01}_{-0.03}$ & - \\
\hline
\end{tabular}
\end{center}
\end{table*}

\begin{table*}[h!]
\begin{center}
\setlength{\extrarowheight}{3pt}
\caption{High-mass transiting planets used in the analysis of radii.}
\label{tab:highplanets}
\begin{tabular}{lrrrrrrrr}
\hline
ID & $M_p$ ($M_J$) & $R_p$ ($R_J$) & $a$ (AU) & $e$ & $T_{eff}$ (K) & $[$Fe/H$]$ & $R_{\ast}$ ($R_{\odot}$) & Age (Gy) \\
\hline
HAT-23b & $2.09\pm0.11$ & $1.37\pm0.09$ & $0.0232\pm0.0002$ & 0 & $5905\pm80$ & $0.15\pm0.04$ & $1.20\pm0.04$ & 4.0 \\
WASP-46b & $2.10\pm0.07$ & $1.31\pm0.05$ & $0.0245\pm0.0003$ & 0 & $5620\pm160$ & $-0.37\pm0.13$ & $0.92\pm0.03$ & 1.4 \\
Kepler-5b & $2.11\pm0.06$ & $1.43\pm0.05$ & $0.0506\pm0.0007$ & 0 & $6297\pm60$ & $0.04\pm0.06$ & $1.79\pm0.05$ & 0.6 \\
HAT-22b & $2.15\pm0.06$ & $1.08\pm0.06$ & $0.0414\pm0.0005$ & 0 & $5302\pm80$ & $0.24\pm0.08$ & $1.04\pm0.044$ & 12.4 \\
HAT-31b & $2.17^{0.11}_{0.08}$ & $1.07^{0.24}_{0.16}$ & $0.0550\pm0.0150$ & 0.25 & $6065\pm100$ & $0.15\pm0.08$ & $1.36^{0.27}_{0.18}$ & 3.17 \\
WASP-8b & $2.23\pm0.17$ & $1.17^{+0.18}_{-0.06}$ & $0.0793\pm0.003$ & 0.31 & $5600\pm80$ & $0.17\pm0.07$ & $0.95^{+0.03}_{-0.06}$ & 1.0 \\
HAT-14b & $2.23\pm0.06$ & $1.15\pm0.05$ & $0.0606\pm0.0007$ & 0.10 & $6600\pm90$ & $0.11\pm0.08$ & $1.47\pm0.05$ & 0.4 \\
WASP-36b & $2.28\pm0.07$ & $1.27\pm0.03$ & $0.0262\pm0.0003$ & 0 & $5881\pm138$ & $-0.31\pm0.12$ & $0.94\pm0.02$ & 3.0 \\
CoRoT-11b & $2.33\pm0.34$ & $1.43\pm0.03$ & $0.044\pm0.005$ & 0 & $6440\pm120$ & $-0.03\pm0.08$ & $1.37\pm0.03$ & 2.0 \\
CoRoT-17b & $2.45\pm0.16$ & $1.02\pm0.07$ & $0.0461\pm0.0008$ & 0 & $5740\pm80$ & $0.0\pm0.1$ & $1.59\pm0.07$ & 10.7 \\
Kepler-17b & $2.45\pm0.01$ & $1.31\pm0.02$ & $0.0259\pm0.0004$ & 0 & $5781\pm85$ & $0.26\pm0.10$ & $1.05\pm0.03$ & 1.78 \\
Qatar-2b & $2.49\pm0.09$ & $1.14\pm0.04$ & $0.0215\pm0.0004$ & 0 & $4645\pm50$ & $0.0\pm0.1$ & $0.71\pm0.02$ & - \\
CoRoT-21b & $2.53\pm0.37$ & $1.30\pm0.14$ & $0.0417\pm0.0011$ & 0 & $6200\pm100$ & $0.0\pm0.1$ & $1.95\pm0.21$ & 4.1 \\
WASP-38b & $2.71\pm0.07$ & $1.08\pm0.05$ & $0.0755^{+0.0008}_{-0.0009}$ & 0.03 & $6150\pm80$ & $-0.12\pm0.07$ & $1.37^{+0.05}_{-0.04}$ & - \\
CoRoT-23b & $2.80\pm0.25$ & $1.05\pm0.13$ & $0.0477\pm0.0038$ & 0.16 & $5900\pm100$ & $0.05\pm0.10$ & $1.61\pm0.18$ & 7.2 \\
CoRoT-6b & $2.96\pm0.34$ & $1.17\pm0.04$ & $0.0855\pm0.0015$ & 0 & $6090\pm70$ & $-0.2\pm0.1$ & $1.03\pm0.03$ & - \\
WASP-10b & $3.06^{+0.23}_{-0.21}$ & $1.08\pm0.02$ & $0.0371^{+0.0014}_{-0.0013}$ & 0 & $4675\pm100$ & $0.03\pm0.2$ & $0.78\pm0.04$ & 0.2 \\
CoRoT-2b & $3.31\pm0.16$ & $1.47\pm0.03$ & $0.0281\pm0.0009$ & 0 & $5625\pm120$ & $0.0\pm0.1$ & $0.90\pm0.02$ & - \\
CoRoT-18b & $3.47\pm0.38$ & $1.31\pm0.18$ & $0.0295\pm0.0016$ & 0 & $5440\pm100$ & $-0.1\pm0.1$ & $1.00\pm0.13$ & 0.6 \\
WASP-33b & $3.50\pm0.6$ & $1.50\pm0.02$ & $0.0256\pm0.0002$ & 0 & $7400\pm200$ & $0.1\pm0.2$ & $1.44\pm0.03$ & - \\
WASP-32b & $3.60\pm0.07$ & $1.18\pm0.07$ & $0.0394\pm0.0003$ & 0 &$6100\pm100$ & $-0.13\pm0.10$ & $1.11\pm0.05$ & - \\
HAT-21b & $4.06\pm0.16$ & $1.02\pm0.09$ & $0.0494\pm0.0007$ & 0.23 & $5588\pm80$ &$0.01\pm0.08$ & $1.11\pm0.08$ & 10.2 \\
HAT-16b & $4.19\pm0.09$ & $1.29\pm0.07$ & $0.0413\pm0.0004$ & 0.04 & $6158\pm80$ & $0.17\pm0.08$ & $1.24\pm0.05$ & 2.0 \\
CoRoT-20b & $4.24\pm0.23$ & $0.84\pm0.04$ & $0.0902\pm0.0021$ & 0.56 & $5880\pm90$ & $0.14\pm0.12$ & $1.02\pm0.05$ &  0.1 \\
OGLE2-L9b & $4.5\pm1.0$ & $1.61\pm0.04$ & $0.0308\pm0.0005$ & 0 & $6933\pm58$ & $-0.05\pm0.2$ & $1.53\pm0.04$ & 0.3 \\
HAT-20b & $7.25\pm0.19$ & $0.87\pm0.03$ & $0.0361\pm0.0005$ & 0 & $4595\pm80$ & $0.35\pm0.08$ & $0.69\pm0.21$ & 6.7 \\
CoRoT-14b & $7.6\pm0.6$ & $1.09\pm0.07$ & $0.0270\pm0.0020$ & 0 & $6035\pm100$ & $0.05\pm0.15$ & $1.21\pm0.08$ & 0.6 \\
WASP-14b & $7.73^{+0.43}_{-0.67}$ & $1.26^{+0.08}_{-0.06}$ & $0.037^{+0.001}_{-0.002}$ & 0.09 & $6475\pm100$ & $0.0\pm0.2$ & $1.30\pm0.05$ & 0.25 \\
Kepler-14b & $8.40\pm0.19$ & $1.14^{0.07}_{0.05}$ & $0.0819^{0.0053}_{0.0041}$ & 0 & $6395\pm60$ & $0.12\pm0.06$ & $2.05^{0.11}_{0.08}$ & 2.2 \\
HAT-2b & $9.09\pm0.24$ & $1.16^{+0.07}_{-0.09}$ & $0.0688\pm0.0007$ & 0.52 & $6290\pm60$ & $0.14\pm0.08$ & $1.64^{+0.09}_{-0.08}$ & 0.5 \\
WASP-18b & $10.43\pm0.40$ & $1.17\pm0.08$ & $0.0205\pm0.0004$ & 0 & $6400\pm100$ & $0.0\pm0.09$ & $1.23\pm0.05$ & 0.53 \\
XO-3b & $11.79\pm0.59$ & $1.22\pm0.07$ & $0.0454\pm0.0008$ & 0.26 & $6429\pm100$ & $-0.177\pm0.08$ & $1.38\pm0.08$ & 0.82 \\
\hline
\end{tabular}
\end{center}
\end{table*}

We used updated parameters for WASP-17b, as given in \citet{anderson11}, with a planetary radius of 2.0~$R_J$ and eccentricity 0.03 found from secondary eclipse timing, for WASP-7b, as given in \citet{southworth11}, with a planetary radius of 1.33~$R_J$, and for CoRoT-13b, with a radius of 1.25~$R_J$ \citep{southworth11b}.

We fixed the eccentricities of many of these planets to zero: WASP-6b, WASP-12b, WASP-22b, WASP-28b and WASP-32b were reported to have eccentricities of 0.05 \citep{gillon09}, 0.02 \citep{hebb09}, 0.02 \citep{maxted10a}, 0.05 \citep{west10} and 0.02 \citep{maxted10b} respectively. However, Lucy-Sweeney tests \citep{lucy71} on the original data give values of 0.22, 0.17, 0.26, 1.0 and 0.68 respectively for the probability that the measured eccentricities could have arisen by chance from underlying circular orbits, given the uncertainties and uneven sampling of the radial-velocity observations, showing that in all these cases a circular orbit is likely.

Secondary eclipse timings often show eccentricities to be considerably lower than the upper limits allowed by radial-velocity measurements alone. For WASP-12b, \citet{husnoo11} and \citet{campo10} find a likely circular orbit, based on secondary eclipse timings. Similarly, WASP-19b was attributed an eccentricity of 0.02 in \citet{hebb10} but \citet{gibson10} find a circular orbit from the timing of the measured secondary eclipse, though conversely, a small eccentricity (0.009) in the orbit of WASP-18b was confirmed to $7\sigma$ by \citet{nymeyer11} from their secondary eclipse measurements. WASP-10b was originally reported to have an eccentricity of 0.06 \citep{christian09}, but \citet{maciejewski10} find that a circular orbit is likely, with the difference in radial velocities due to starspot activity.

Additionally, \citet{pont11} point out that a finite best-fit eccentricity will always be found where the eccentricity is allowed to float as a free parameter in fitting the data. Spurious eccentricities with apparent significance as great as 3$\sigma$ often arise from a combination of measurement error and uneven sampling around the orbit. From a homogeneous re-analysis of available data, \citet{pont11} find that the eccentricities of CoRoT-5b and HAT-13b should be set to zero. For other planets in our sample announced with small eccentricity values, we also fix them to zero if the values are within 3$\sigma$ of 0, which was the case for HAT-18b, -19b, -20b, -22b, -23b, -25b, -28b, -29b, -32b and -33b, CoRoT-12b, -18b and -19b, Kepler-12b, -14b and -17b, WASP-23b and -50b and HD~209458b, for which the data has previously been noted to be consistent with a circular orbit \citep{laughlin05, snellen08}.

This resulted in 17 out of 119 planets having an eccentricity greater than zero, as shown in Tables \ref{tab:satplanets} to \ref{tab:highplanets}, with significantly more high-mass planets (greater than 2.0~$M_J$) showing a non-zero eccentricity: 11 of 32 high-mass planets, compared to 1 of 16 Saturn-mass (0.1-0.5~$M_J$) and 4 of 71 Jupiter-mass (0.5-2.0~$M_J$) planets. This difference does not seem to be explained by the known period-eccentricity relationship \citep[for example]{halbwachs05}, that exoplanets on short periods have circular orbits due to tidal circularisation while exoplanets on wider orbits have a range of eccentricity values, since both low-mass and high-mass eccentric and non-eccentric planets cover a range of orbital periods. The difference may therefore lend support to the trend discussed by \citet{ribas07} for radial velocity-discovered planets that high-mass planets have an eccentricity distribution closer to binary stars than low-mass planets. It should be noted though that it is easier to detect genuine eccentricity in high-mass planets than in low-mass ones because the velocity measurement errors are small relative to the total RV amplitude.

To determine the factors that have a significant effect on planetary radii, we performed multivariate regression analyses (Singular Value Decomposition (SVD) fits), weighted by the uncertainties on the measurements for planetary radii, for a variety of explanatory parameters, including the planetary equilibrium temperature, $T_{eq}$, in Kelvin, semi-major axis, $a$, in AU, planetary mass, $M$, in Jupiter masses, stellar metallicity, $[$Fe/H$]$ and tidal heating rate on the planet, $H_{tidal}$, divided by $1\times10^{20}$W. The equilibrium temperature of the planet is given by
\begin{equation}
T_{eq} = T_{\ast,eff} \left( \frac{1-A}{F} \right)^{\frac{1}{4}} \sqrt{\frac{ R_{\ast}}{2 a}}
\end{equation}
\noindent where $T_{\ast,\mbox{eff}}$ is the host star's effective temperature, $R_{\ast}$ is the stellar radius, $a$ is the semi-major axis, $A$ is the planetary Bond albedo and $F$ is the fraction of the planet surface that is re-radiating flux.  Albedo values have not so far been determined for the majority of exoplanets, but \citet{rowe08} determined a very low albedo for HD~209458b, finding just $0.04\pm0.05$ for the geometric albedo and inferring a Bond albedo of less than 0.12 from this. We set $A=0$ ($F=1$) here to calculate the equilibrium temperatures. For a given $A$ and $F$ then, $T_{eq}$ is directly proportional to the irradiating flux from the star. 

The tidal heating rate is given by
\begin{equation}
H_{tidal} = \frac{63}{4} \frac{(G M_{\ast})^{3/2} M_{\ast} R_p^5 a^{-15/2} e^2}{3Q / 2k}
\end{equation}
\noindent where $Q$ is the tidal dissipation factor (taken to be $10^5$) and $k$ is the Love number (=0.51) \citep{jackson08},


To determine the best-fitting SVD model, we used the Bayesian Information Criterion, BIC, value where
\begin{equation}
BIC = \chi^2 + N_P \ln(N_D)
\end{equation}
\noindent where $N_P$ is the number of parameters used in the fit and $N_D$ is the number of datapoints \citep{szydlowski08}. Initially, we used a one-parameter model, and used the BIC values to determine the parameter with the strongest influence on planetary radii. Next, we tried a two-parameter model, using the best parameter from the one-parameter fits, and adding each of the other parameters in turn. If a two-parameter model was found to have a lower BIC value than the best one-parameter model, we selected this and attempted a three-parameter model using the best two-parameter model plus a further parameter, and so on until the best model was found. 

\subsection{All planets}

\begin{figure}[h!]
\includegraphics[angle=90,width=3.6in]{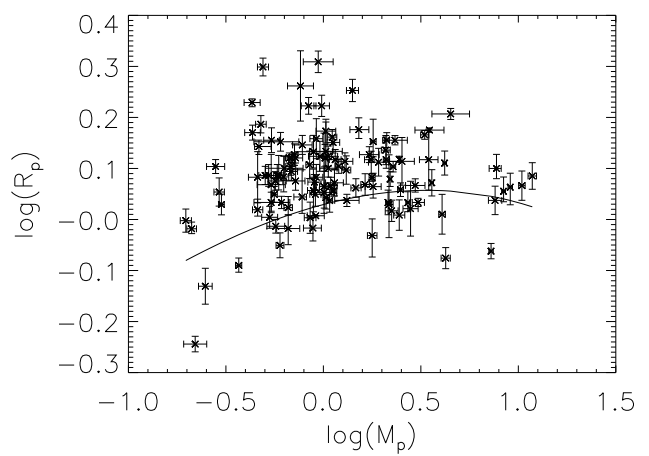}
\caption{Mass-Radius plot of 119 transiting exoplanets showing the expected mass-radius relationship for pure H (solid line) \citet{seager07}}
\label{fig:mrplotsing}
\end{figure}
\begin{figure}[h!]
\includegraphics[angle=90,width=3.6in]{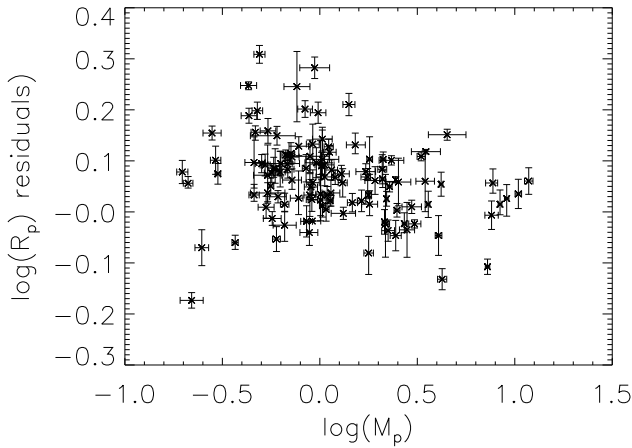}
\caption{Mass against radius residuals after subtracting the \citet{seager07} mass-radius relationship.}
\label{fig:mresplotsing}
\end{figure}

To find the best-fitting model for planetary radius for all 119 planets over the full mass range of 0.1 to 12.0~$M_J$, we initially subtracted the expected cold-body mass-radius relationship determined by \citet{seager07} using interior models. Figure \ref{fig:mrplotsing} shows log$(M/M_J)$ against log$(R/R_J)$ for the 119 planets used in this analysis, with the \citet{seager07} mass-radius relationship for pure hydrogen with a solid line. Most of the planets follow the pure H relationship, or lie above it, while a few (notably CoRoT-8b), perhaps strongly core-dominated, lie below it. We subtracted the logarithmic mass-radius relationship expected for pure H from the planets, leaving the log($M/M_J$)-log($R/R_J$) scatter shown in \ref{fig:mresplotsing}.


We then performed an SVD analysis on the radius residuals, and found a best-fit model where
\begin{eqnarray}
\mbox{log}(R_p/R_J) &=& -1.374 + 0.612 \mbox{log}(T_{eq}/K) \nonumber \\
 &+& 0.364 \mbox{log}(a/AU)  \nonumber \\
 &+& 0.043 \mbox{log}(H_{tidal}/1\times10^{20} \nonumber \\
 &-& 0.023 [\mbox{Fe/H}]
\end{eqnarray}
\noindent A table of BIC values summarising the model building process is given in Table \ref{tab:allbic}.


The use of the logarithm of planetary equilibrium temperature may incorporate several possible heating effects showing a power-law relationship of planetary temperature with radius.

A relationship between planetary radius was not only found with planetary equilibrium temperature, as expected, but also with the semi-major axis. The relationship with temperature contains a separate and opposing dependence on semi-major axis than that seen in the direct relationship with semi-major axis where the planetary radius tends to be smaller for closer-in planets. This tendency can actually be seen directly by comparing pairs of planets for which parameters other than the semi-major axis are very similar. For example WASP-19b and WASP-4b have very similar mass (1.15 and 1.12~$M_J$, respectively), metallicity ($\sim$0) and equilibrium temperatures (1995 and 1875~K, respectively) but WASP-19b has a radius of $1.31 R_J$ at 0.016~AU while WASP-4b has $1.42 R_J$ at 0.023~AU. Direct comparisons between planets are generally difficult due to the different values of the various parameters affecting the radii, and thus analyses such as set out here can help to disentangle each effect. The direct relationship of semi-major axis on radius is seen more clearly when analysing sub-samples of planets, below, and is discussed further in those sections.

\begin{figure}[h!]
\includegraphics[angle=90,width=3.6in]{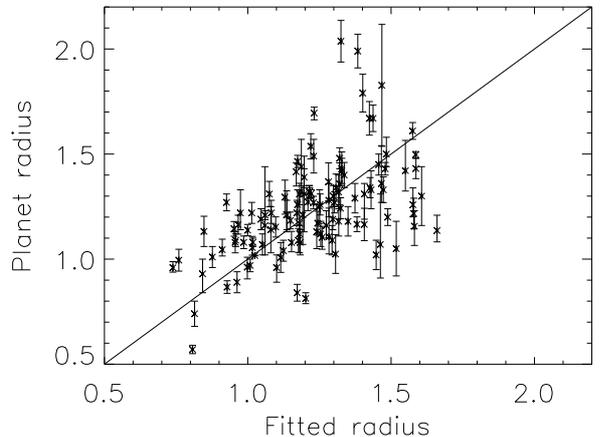}
\caption{Results of the radius calibration on all 119 transiting exoplanets using the \citet{seager07} mass-radius relationship and a single four-parameter SVD fit.}
\label{fig:allsvd}
\end{figure}

A plot of observed planetary radii against fitted radii using the above equation is given in Figure \ref{fig:allsvd}, showing significant scatter (mean error in fitted radius of 0.16~$R_J$), with 35 of 119 fitted radii values more than 0.2~$R_J$ from the observed values (7 with more than 0.4~$R_J$ error). This calibration was not therefore felt to provide a very good fit to the data. We then partitioned the 119 planets into 16 Saturn-mass planets, $0.1-0.5M_J$, dealt with in Section \ref{sub:saturns}, 71 Jupiter-mass planets, $0.5-2.0M_J$, discussed in Section \ref{sub:jupiters}, and 32 high-mass planets, $2.0-12.0M_J$, in Section \ref{sub:high}. For each subset of planets, we performed SVD fits as above, without subtracting the \citep{seager07} mass-radius relationships first. We discuss the significant effects on radii for each mass range below, and provide a best-fitting equation.

\begin{table*}[h!]
\begin{center}
\setlength{\extrarowheight}{5pt}
\caption{BIC values for SVD fits on all 119 transiting exoplanets.}
\label{tab:allbic}
\begin{tabular}{lrrrr}
\hline
\# parameters & log($T_{eq}$) & log($a$) & $[$Fe/H$]$ & log($H_{tidal}$) \\
\hline
1 & 2789 & 4669 & 4614 & 4735 \\
2: $T_{eq} +$ & - & 2132 & 2796 & 2791 \\
3: $T_{eq} + a + $ & - & - & 2103 & 2070 \\
4:  $T_{eq} + a +$ [Fe/H] + & - & - & 2042 & - \\
\hline
\end{tabular}
\end{center}
\end{table*}

\subsection{Saturn-mass planets}
\label{sub:saturns}

Sixteen known transiting exoplanets fall into the mass range 0.1-0.5$M_J$ (including WASP-6b). The upper mass cutoff of $0.5 M_J$ was chosen to represent the point where the mass-radius relationship changes slope from a steep correlation in the low mass planets to an almost flat relationship in the Jupiter-mass group, thereby splitting the planets into distinct groups with different relationships between radius and mass.

The best-fitting model for this mass range was found using the procedure described for the global SVD fit above, with the model BIC values provided in Table \ref{tab:lowbic}, and was found to be a four-parameter model, mainly determined by planet mass and host star metallicity:
\begin{eqnarray}
\mbox{log}(R_p/R_J) &=& -0.077 + 0.450 \mbox{log}(M_p/M_J) \nonumber \\
 &-& 0.314 [\mbox{Fe/H}] + 0.671 \mbox{log}(a/AU) \nonumber \\
 &+& 0.398 \mbox{log}(T_{eq}/K)
\label{eq:sat}
\end{eqnarray}
\noindent The equilibrium temperature of the planet only minimally affects the planet radius here (though it has a strong effect on more massive planets, below). We did not include tidal heating in the fit, since only 1 of the 16 planets have eccentricities demonstrably greater than zero (WASP-17b).

Eight planets have radii with residuals greater than $\pm$0.2~$R_J$ relative to the fitted model: HAT-18b (0.20), HAT-19b (0.31) and Kepler-12b (0.26) all have their radii underestimated, such that their observed radii are greater than expected from the model fit. The radii of CoRoT-5b, CoRoT-8b, WASP-6b, WASP-13b and WASP-21b are overestimated by 0.32, 0.21, 0.20, 0.26 and 0.47~$R_J$ respectively.

\begin{table*}[h!]
\begin{center}
\setlength{\extrarowheight}{5pt}
\caption{BIC values for SVD fits on 16 low-mass transiting exoplanets.}
\label{tab:lowbic}
\begin{tabular}{lrrrr}
\hline
\# parameters & log($T_{eq}$) & log($a$)  & log($M$) & $[$Fe/H$]$ \\
\hline
1: & 1396 & 2355 & 900 & 1551 \\
2: $M$ + & 925 & 895 & - & 612 \\
3: $M$ + $[$Fe/H$]$ + & 566 & 531 & - & - \\
4: $M$ + $[$Fe/H$]$ + $a$ + & 470 & - & - & - \\
\hline
\end{tabular}
\end{center}
\end{table*}


\paragraph{Planetary mass}

\begin{figure}[h!]
\includegraphics[angle=0,width=90mm]{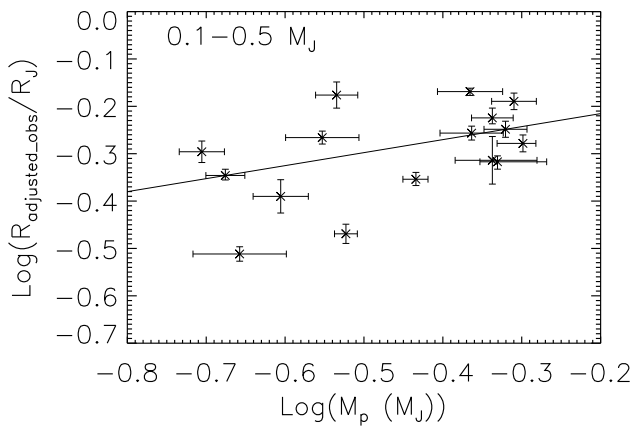}
\caption{The correlation between observed radius and mass for Saturn-mass planets, after removing the effects of the other parameters used in the SVD fit.}
\label{fig:corrm}
\end{figure}

As mentioned in Section \ref{s:intro}, the radii of gaseous planets do not generally depend on their masses but there is an effect on radii for lower-mass planets, made up of essentially incompressible matter. For the Saturn-mass subset of planets, the mass has the strongest effect on planet radius. A correlation between mass and adjusted radius, after removing the effects of the other parameters, produces a coefficient of 0.51, shown in Figure \ref{fig:corrm}, with a least-absolute-deviation model overplotted. The positive coefficients in the correlation and the SVD model fit shows that for low-mass planets, radius increases as mass increases, as expected. 

The fit coefficient for mass, at 0.45, is in fact higher than expected from the polytropic relationship with $n \sim 1$ for incompressible matter, such that $R \propto M^{1/3}$, or log~$R \propto 0.33$~log~$M$. The additional dependence of radius on mass indicates that the density of the planet itself is dependent on mass, suggesting that the core-envelope ratio increases towards lower mass, as recently shown in \citet{miller11}.

\paragraph{Metallicity}

\begin{figure}[h!]
\includegraphics[angle=0,width=90mm]{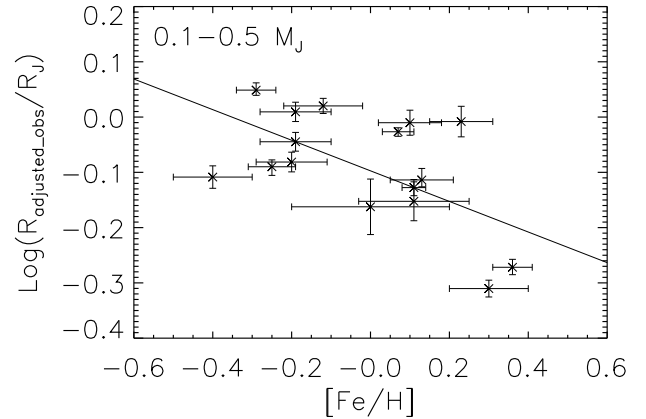}
\caption{The correlation between observed radius and metallicity for Saturn-mass planets, after removing the effects of the other parameters used in the SVD fit.}
\label{fig:corrfe}
\end{figure}

A correlation between the adjusted radius, after removing effects due to other parameters, and the host star metallicity results in a coefficient of -0.56, shown in Figure \ref{fig:corrfe}. The strong negative coefficient, as well as the negative coefficient value in the SVD fit, indicates that as the metallicity of the host star increases, planetary radius decreases. This lends support to the theory that higher system metallicity produces larger planetary cores, and thus smaller radii \citep{guillot06}. Given that a strong correlation is found at all between host star metallicity and planetary radius, this implies that a planet's heavy element mass fraction is related to that of its host star.


\paragraph{Planet-star separation}

\begin{figure}[h!]
\includegraphics[angle=0,width=90mm]{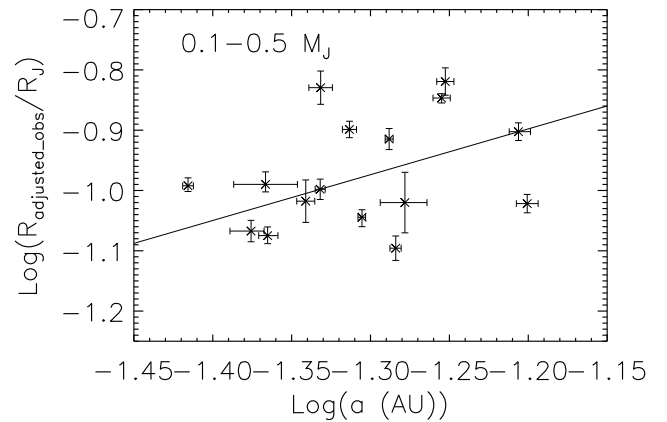}
\caption{The correlation between observed radius and planet-star distance for Saturn-mass planets, after removing the effects of the other parameters used in the SVD fit.}
\label{fig:corra}
\end{figure}

A small effect on BIC value was also produced by the star-planet separation, however a correlation between the semi-major axis and the observed radius, after removing the effects on radius of the other terms by subtracting these parameters multiplied by their coefficients above, produces a low coefficient value of 0.34, shown in Figure \ref{fig:corra}, clearly not a strong effect in this mass range. 


\paragraph{Stellar irradiation} 

\begin{figure}[h!]
\includegraphics[angle=0,width=90mm]{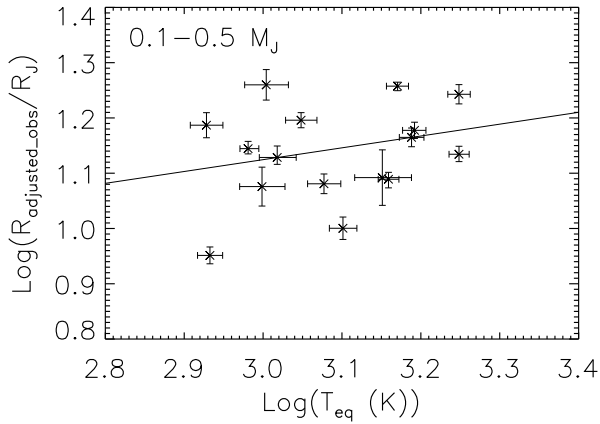}
\caption{The correlation between observed radius and equilibrium temperature for Saturn-mass planets, after removing the effects of the other parameters used in the SVD fit.}
\label{fig:corrteq}
\end{figure}

The equilibrium temperature of the planet also has an effect on the radii. Performing a simple correlation between the logarithms of observed radius and equilibrium temperature, after adjusting for the other terms in the radius fit, gives a weak coefficient of 0.28, shown in Figure \ref{fig:corrteq}. The positive coefficient shows that planets of a higher equilibrium temperature generally have larger radii, but for core-dominated Saturn-mass planets, this appears to be only a minor effect. The apparent contrast in results found here in comparison to \citet{enoch10} is due to lowering the upper mass cutoff from $0.6 M_J$ to $0.5 M_J$ here. In \citet{enoch10}, over half the planets in the small sample analysed were in the mass range 0.5-0.6~$M_J$, which is here part of the larger Jupiter-mass planet group where a large effect of equilibrium temperature on radius is seen (see below).
 
\paragraph{Age}

All 16 planets have published values for isochrone age. To see if the age of a system affects the radius of close-in planets, for example due to Kelvin-Helmholtz contraction, we tried adding a parameter for age to the model, but it produced no improvement in the model, with BIC = 474 compared to 470 without the age parameter. The expected Kelvin-Helmholtz contraction is likely to be mostly offset by the continued heating from strong stellar irradiation on the close-in planets.

\subsection{Jupiter-mass planets}
\label{sub:jupiters}

This is the largest subset of transiting planets, with 71 having a mass between 0.5 and 2.0~$M_J$. A fit to all 71 planets resulted in a two-parameter fit given by
\begin{eqnarray}
\mbox{log}(R_p/R_J) &=& -2.217 + 0.856 \mbox{log}(T_{eq}/K) \nonumber \\
   &+& 0.291 \mbox{log}(a/AU) \nonumber \\
 \end{eqnarray}
\noindent Again, we did not attempt to fit tidal heating, since only 4 of the 71 planets have non-zero eccentricity. This model produced a mean error of 0.106~$R_J$ in fitted radii, with five planets having errors in fitted radii of greater than $0.2 R_J$ (up to $0.27 R_J$), plus one significant outlier, HAT-32b, underestimated by 0.63~$R_J$. HAT-32b is an extremely bloated planet with a measured radius of over 2~$R_J$, which may be in a slightly eccentric orbit \citep{hartman10}. Tidal heating on the planet could potentially explain the extra inflation above that fitted here. 

The addition of a coefficient for mass produced a small improvement of 20.1 in BIC value, but increased the mean error in fitted radii to 0.113, so was felt to be insignificant, particularly since a correlation of mass with radius adjusted for the fitted effects of temperature and semi-major axis yielded a very small coefficient of 0.13. Model BIC values are given in Table \ref{tab:midbic}.



\begin{table*}[h!]
\begin{center}
\setlength{\extrarowheight}{5pt}
\caption{BIC values for SVD fits on 71 Jupiter-mass transiting exoplanets.}
\label{tab:midbic}
\begin{tabular}{lrrrrr}\hline
\# parameters & log($T_{eq}$) & log($a$) & log($M$) & $[$Fe/H$]$ \\
\hline
1 & 797 & 1287 & 1628 & 1436 \\
2: $T_{eq}$ + & - & 402 & 765 & 406 \\
3: $T_{eq} + a$ + & - & - & 382 & 406 \\
\hline
\end{tabular}
\end{center}
\end{table*}

Planetary temperature is the strongest parameter in the fit: the high planetary equilibrium temperatures sustained due to the strong stellar irradiation on close-in exoplanets produces a shallow temperature gradient in the planet, impeding the interior cooling and thus contraction. These planets must have migrated early and not contracted as a planet orbiting further from the star would have \citep{guillot02} since if they had formed and cooled at a larger semi-major axis before inward migration, their radii would only be marginally increased due to a puffed-up atmosphere \citep{iro05,burrows00} and such a strong effect on radius due to irradiation would not be seen.

\begin{figure}[h!]
\includegraphics[angle=0,width=90mm]{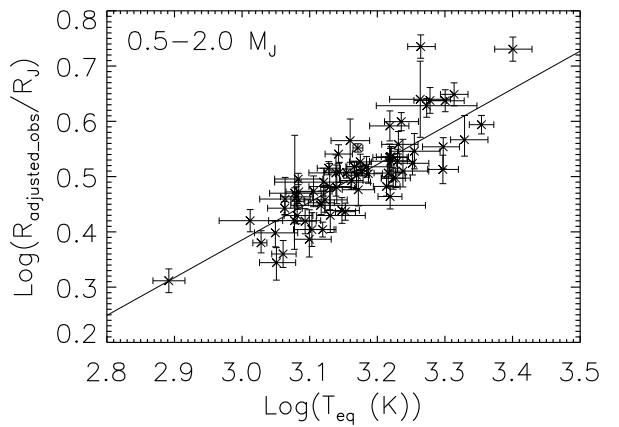}
\caption{The correlation between observed radius and equilibrium temperature for Jupiter-mass planets, after removing the effects of semi-major axis.}
\label{fig:corrteqmid}
\end{figure}

\begin{figure}[h!]
\includegraphics[angle=0,width=90mm]{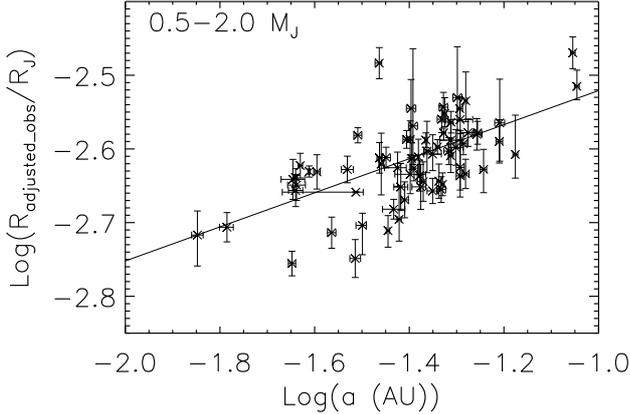}
\caption{The correlation between observed radius and semi-major axis for Jupiter-mass planets, after removing the effects of equilibrium temperature.}
\label{fig:corramid}
\end{figure}

Correlations of radius with equilibrium temperature and semi-major axis, after removing effects on the radius due to the other terms, give coefficients of 0.84 and 0.62, respectively, shown in Figures \ref{fig:corrteqmid} and \ref{fig:corramid}. The positive correlation and fit coefficient for the semi-major axis indicate that the closer the planet is to the star, the smaller the radius. This could be due to the ease of atmospheric escape described in Section \ref{s:intro}, where the atmospheric escape of closer planets is enhanced by the $K_{tide}$ factor given in Equation \ref{ktide}. As mentioned above, WASP-19b has a $K_{tide}$ value of 0.29 at 0.016 AU, and its measured radius is 0.11~$R_J$ lower than that of WASP-4b, of similar mass, equilibrium temperature and host star metallicity, orbiting slightly farther from its host star.

In this mass range, planet mass has no apparent effect on planet radius. This is as expected from the relationship $R \propto M^{(1-n)/(3-n)}$, as discussed in Section \ref{s:intro}, where the polytropic index is $n\sim1$ for Jupiter-mass planets. A lack of mass-radius effect in this mass range also indicates that, in contrast to the Saturn-mass planets discussed above, the cores of these planets are either unimportant, or are not correlated with planet mass. 

Adding a coefficient for age to the model increases the BIC value from 218 to 220, for the 57 planets with published isochrone ages. To investigate whether the relationship between semi-major axis and radius has any correlation with the isochrone age of the system, we altered the model parameter from log($a$) to log($a$)/log($age$). This increased the model BIC value to 306. If there is no dependence here on age, a reduction in radius of close-in exoplanets due to blow-off could occur early in the system history. The T Tauri phase \citep{joy45,herbig52} in the evolution of low-mass stars ($< 2 M_{\odot}$) produces strong XUV radiation and intense stellar winds that cause the disappearance of the gas component of the disc \citep{hayashi81} via outflow. The $K_{tide}$ enhanced atmospheric escape due to the XUV radiation combined with the strong stellar wind could strip these close exoplanets of their outer layers of atmosphere, reducing their radii. \citet{murrayclay09} calculated that a hot Jupiter at 0.05~AU from its host star could lose around $6\times10^{12}$g s$^{-1}$ due to the intense XUV from a T-Tauri star, equating to around 1\% of a Jupiter mass during the $\sim$100~Myr T Tauri phase; more for a planet orbiting closer than 0.05~AU. Thus the relationships of radius with semi-major axis and equilibrium temperature could both be due to proximity of the planet to the host star, but from entirely different processes, namely atmospheric blow-off and retarded contraction, respectively. However, a small reduction in the overall mass of a Jupiter-mass planet is not likely to produce a noticeable effect on radius in this mass regime, so the radius-semi-major axis relationship may be due to another factor, perhaps a migration halting mechanism resulting in planets with larger cores / higher heavy element content orbiting closer to their host stars than otherwise similar planets with lower heavy element content.

\subsection{High-mass planets}
\label{sub:high}

Thirty-two transiting exoplanets used in this analysis have a mass in the range 2.0-12.0~$M_J$. Equilibrium temperature, planet mass, metallicity and tidal heating were found to be important in determining their radii. The SVD fit resulted in 
\begin{eqnarray}
\mbox{log}(R_p/R_J) &=& -1.067 + 0.380 \mbox{log}(T_{eq}/K) \nonumber \\
  &-& 0.093 \mbox{log}(M_p/M_J) - 0.057 [\mbox{Fe/H}] \nonumber \\
     &+& 0.019 \mbox{log}(H_{tidal}/1\times10^{20}) 
\end{eqnarray}
\noindent with model BIC values provided in Table \ref{tab:highbic}. The mean error in fitted radius is 0.10~$R_J$, with four planets having an error in fitted radius greater than 0.2~$R_J$ (WASP-18b, CoRoT-2b, CoRoT-17b and CoRoT-23b), though none was greater than 0.3~$R_J$. Adding a coefficient for semi-major axis did not improve the BIC.


\begin{table*}[h!]
\begin{center}
\setlength{\extrarowheight}{5pt}
\caption{BIC values for SVD fits on 32 high-mass transiting exoplanets.}
\label{tab:highbic}
\begin{tabular}{lrrrrr}
\hline
\# parameters & log($T_{eq}$) & log($a$) & log($M$) & $[$Fe/H$]$ & log($H_{tidal})$ \\
\hline
1 & 239 & 679 & 911 & 931 & 864 \\
2: $T_{eq}$ + & - & 243 & 201 & 222 & 241 \\
3: $T_{eq} + M$ + & - & 205 & - & 192 & 193 \\
4: $T_{eq} + M + [$Fe/H$]$ + & - & 196 & - & - & 185 \\
5: $T_{eq} + M + [$Fe/H$] + H_{tidal}$ + & - & 187 & - & - & - \\
\hline
\end{tabular}
\end{center}
\end{table*}

\begin{figure}[h!]
\includegraphics[angle=0,width=90mm]{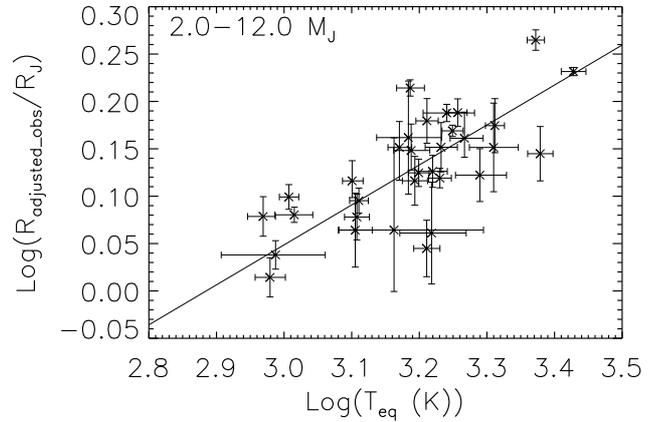}
\caption{The correlation between observed radius and equilibrium temperature for high-mass planets.}
\label{fig:corrteqhigh}
\end{figure}

The correlation of equilibrium temperature and high-mass planet radii gives a coefficient of 0.72, shown in Figure \ref{fig:corrteqhigh}.

\begin{figure}[h!]
\includegraphics[angle=0,width=90mm]{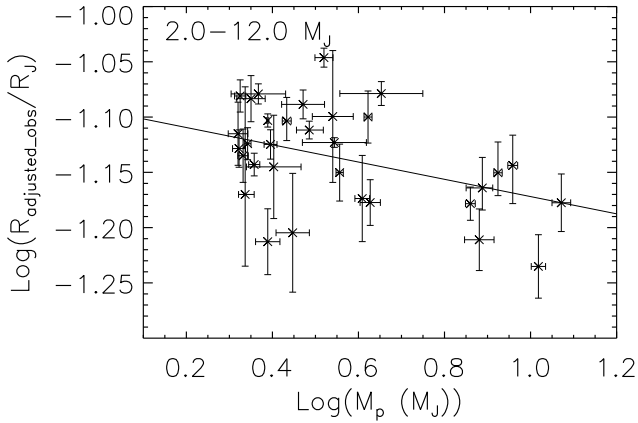}
\caption{The correlation between observed radius and mass for high-mass planets, after removing the effects of other terms.}
\label{fig:corrmhigh}
\end{figure}

As expected, the mass term again becomes important for more massive planets, greater than a few Jupiter masses, where electron degeneracy in the dense cores produces a contraction of the planet and thus a smaller radius. The correlation of planet mass with radius adjusted for equilibrium temperature and semi-major axis effects is -0.47, as seen in Figure \ref{fig:corrmhigh}. Planet mass here thus has the opposite effect on radius as it does on low-mass planets, where the correlation and SVD fit coefficient were positive. 

The addition of a term for tidal heating also has a small effect on the fit - 12 of the 32 planets have non-zero eccentricity. A correlation between tidal heating and adjusted radius gives a coefficient of 0.28.

\begin{figure}[h!]
\includegraphics[angle=0,width=90mm]{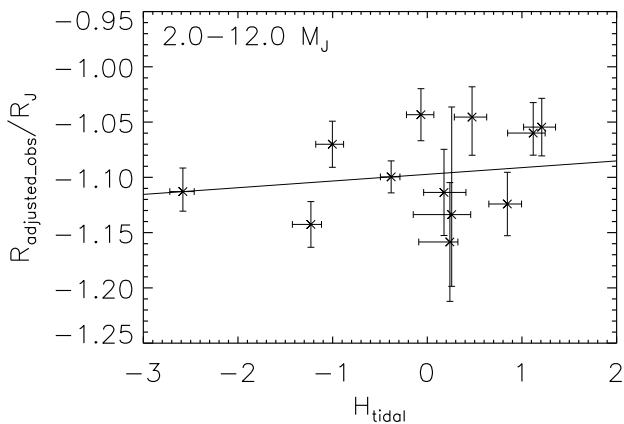}
\caption{The correlation between observed radius and tidal heating for high-mass planets of non-zero eccentricity, after removing the effects of the other terms.}
\label{fig:corrheathigh}
\end{figure}

\begin{figure}[h!]
\includegraphics[angle=0,width=90mm]{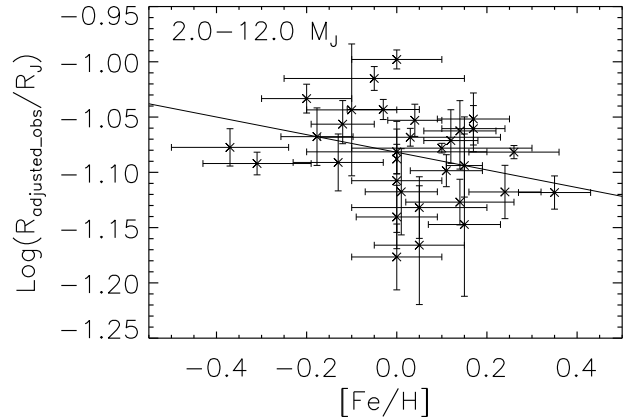}
\caption{The correlation between observed radius and metallicity for high-mass planets, after removing the effects of the other terms.}
\label{fig:corrfehigh}
\end{figure}

Metallicity also has an effect on planet radius: a radius-metallicity correlation gives a fairly weak coefficient of -0.25, shown in Figure \ref{fig:corrfehigh}, but the inclusion of a term for metallicity does improve the SVD fit.
 

The semi-major axis is no longer seen to have the same effect on planet radii as was seen for lower-mass planets. The higher mass of these planets may act to prevent the atmospheric escape that may produce the correlation of radius with semi-major axis seen in lower-mass planets.

Adding age to the model slightly decreases the model BIC, from 105 to 96, for the 26 planets that have published isochrone ages, so a small age dependence is seen for high-mass planets. This SVD fit produces a small negative coefficient for age of -0.03, showing that radius decreases as the planet ages, as would generally be expected due to Kelvin-Helmholtz contraction.

\section{Discussion}
\label{s:disc}

We have attempted to determine the factors that have an effect on planetary radii, along with the magnitude and direction of each significant effect. A positive relationship between planetary equilibrium temperature and radius is seen for Jupiter and high-mass planets, showing that the higher the equilibrium temperature, the larger the planet radius, though not significantly for the Saturn mass planets. The radius dependence found here is not as strong as that reported recently by \citet{laughlin11} who found a proportionality of $R_p \propto T_{eq}^{1.4}$ above that expected from theoretical calculations. Here, the Saturn-mass and low-mass planets show around $R_p \propto T_{eq}^{0.4}$ while the Jupiter-mass planets show $R_p \propto T_{eq}^{0.9}$, in total. These lower values of proportionality suggest that kinetic heating \citep{guillot02}, expected to lead to $R_p \propto T_{eq}^{0.67}$ \citep{laughlin11}, could be the major source of providing the additional heating to inflate planetary radii, since Ohmic heating should yield a higher dependence on temperature of around $R_p \propto T_{eq}^{2.4}$ \citep{laughlin11}.

A negative relationship between host star metallicity and planetary radius is seen for the Saturn-mass and high-mass planets, showing that as host star metallicity increases, planet radius decreases. This implies firstly that there is a relationship between host star metallicity and the fraction of heavy elements present in a planet, and secondly that the higher metallicity likely results in larger planetary cores, producing smaller radii, both as discussed by \citep{guillot06}. No relationship with metallicity is seen for the gaseous envelope-dominated Jupiter mass planets.

A strong positive relationship between semi-major axis and planetary radius is seen for the Jupiter-mass planets, such that planets closer to their host stars have smaller radii, an effect in opposition to the bloating caused by strong stellar irradiation which tends to make planets closer to their stars larger. This may be due to atmospheric blow-off in the early stages of the system development or a migration stopping mechanism that leads to planets of higher heavy element contect orbiting closer to the star. The relationship is much weaker for the core-dominated Saturn-mass planets, and no such relationship is seen for the high-mass planets, perhaps because their larger gravitational pull prevents atmospheric escape. 

The radii of the planets rises with mass for low-mass planets, is unaffected by mass for Jupiter-mass planets, and falls with mass for the high-mass planets. This changing relationship is as expected with the move from incompressible matter to partially electron degenerate bodies at high mass, though the relationship with mass for Saturn-mass planets is stronger than expected from the polytropic relationship, indicating an additional dependence that may be due to the core-envelope ratio increasing towards lower mass.
 
A clear contrast between the core-dominated Saturn-mass planets of $<0.5 M_J$ and the gaseous envelope-dominated Jupiter-mass planets of 0.5-2.0~$M_J$ is therefore seen in the fits. Planetary mass and heavy element content almost exclusively determine the radius of a Saturn-mass planets, with stellar irradiation having little effect, while stellar irradiation and semi-major axis determine the radius of a Jupiter-mass planet almost entirely.

Once we had determined the best SVD model for each subset of planets, we performed a Monte Carlo SVD analysis of 50,000 runs to determine the uncertainties on the fit coefficients. In each run, each input parameter to the SVD fit was perturbed randomly on a normal distribution about their observed values with standard deviations as their published parameter uncertainty values. A summary of the radius calibration terms and uncertainties is provided in Table \ref{tab:coefs}. 

\begin{figure}[h!]
\includegraphics[angle=90,width=3.6in]{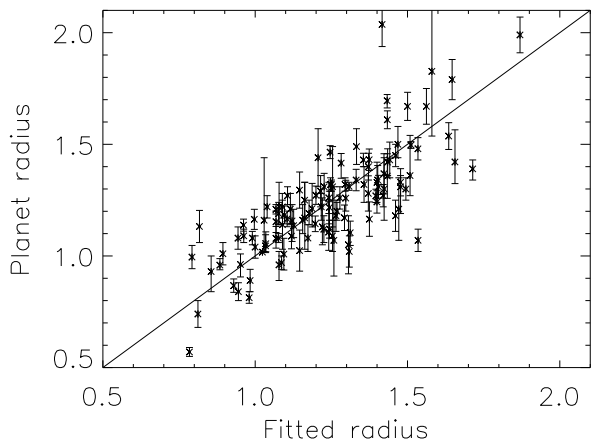}
\caption{Results of the radius calibration on 119 transiting exoplanets using the 3 equations given for each subset.}
\label{fig:rpresult}
\end{figure}

Using the fits obtained for the three subsets of transiting planets to obtain fitted radii for the 119 planets results in Figure \ref{fig:rpresult}, with a generally good agreement between fitted and observed radii. The mean error in fitted radius to the observed radius is 0.11~$R_J$ compared to the mean uncertainty in reported observed radius of 0.07~$R_J$. This solution using three subsets of planets provides a better calibration to radius than the single global analysis, where 35 planets had fitted radius errors of greater than 0.2~$R_J$ (17 greater than 0.3~$R_J$), and the mean error in fitted radius was 0.16~$R_J$. Here, 18 of the 119 planets have errors greater than 0.2~$R_J$ (only 4 greater than 0.3~$R_J$): 8 Saturn-mass planets, 6 Jupiter-mass planets and 4 high-mass planets. The two significant outliers are WASP-21b, with an overestimated fitted radius of 1.54~$R_J$ compared to its measured radius of only 1.07~$R_J$, and HAT-32b with an extremely large measured radius of 2.04~$R_J$ compared to the low fitted radius of 1.41~$R_J$. The simiarly bloated Saturn-mass planet WASP-17b, with measured radius 1.99~$R_J$ is fitted fairly well to 1.87~$R_J$. 


Approximations such as setting the albedo of all planets to zero in calculating planetary equilibrium temperature may have contributed to the poorer radii fits for some planets, and the model scatter may be reduced if all albedos were known. Additionally, the planetary metallicity may sometimes be enhanced or depleted compared to an average planet orbiting a host star of a certain metallicity, which may lead to larger or smaller than expected cores and thus smaller or larger than expected radii.

\begin{table*}[h!]
\begin{center}
\caption{Summary of radius calibration coefficients.}
\label{tab:coefs}
\begin{tabular}{rrrrrrr}
\hline
Mass range ($M_J$) & Constant & $\mbox{log}(T_{eq})$ & $\mbox{log}(a)$ & $\mbox{log}(M_p)$ & $[\mbox{Fe/H}]$ & $\mbox{log}(H_{tidal})$  \\ 
\hline
0.1-0.5 & $-0.077\pm0.698$ & $0.398\pm0.201$ & $0.671\pm0.142$ & $0.450\pm0.153$ & $-0.314\pm0.048$ & - \\
0.5-2.0 & $-2.217\pm0.834$ & $0.856\pm0.284$ & $0.291\pm0.126$ & - & - & - \\
2.0-12.0 & $-1.067\pm0.117$ & $0.380\pm0.036$ & - & $-0.093\pm0.023$ & $-0.057\pm0.028$ & $0.019\pm0.005$ \\
\hline
\end{tabular}
\end{center}
\end{table*}

\bibliographystyle{aa}
\bibliography{rpcal.bib}

\end{document}